\documentclass[journal]{IEEEtran}
\usepackage[utf8]{inputenc}
\usepackage{cite}
\usepackage[cmex10]{amsmath}
\usepackage{amsmath}
\usepackage{amsfonts}
\usepackage{amsthm}
\usepackage{algorithm}
\usepackage{multirow}
\usepackage{multicol}
\usepackage{stfloats}
\usepackage{multicol,multienum}
\usepackage{multirow}
\usepackage[caption=false,font=footnotesize]{subfig}
\usepackage{wrapfig}
\usepackage{color}
\usepackage{graphicx}
\usepackage{xspace}
\usepackage{epstopdf}
\usepackage{suffix}
\usepackage{mathtools}
\usepackage{bm}
\usepackage{hyperref}

\DeclarePairedDelimiterX\MeijerM[3]{\lparen}{\rparen}%
{\begin{smallmatrix}#1 \\ #2\end{smallmatrix}\delimsize\vert\,#3}

\newcommand\MeijerG[8][]{%
  G^{\,#2,#3}_{#4,#5}\MeijerM[#1]{#6}{#7}{#8}}

\WithSuffix\newcommand\MeijerG*[7]{%
  G^{\,#1,#2}_{#3,#4}\MeijerM*{#5}{#6}{#7}}

\begin{document}
\title{Uncertainty Quantification of Multi-Scale Resilience in Nonlinear Complex Networks using Arbitrary Polynomial Chaos}

\author{Mengbang Zou, Luca Zanotti Fragonara,
        Weisi Guo,~\IEEEmembership{Senior Member,~IEEE}
\thanks{M. Zou and L. Zanotti Fragonara are with Cranfield University, Cranfield MK43 0AL, U.K.}
\thanks{W. Guo is with Cranfield University, Cranfield MK43 0AL, U.K., and also with the Alan Turing Institute, London, NW1 2DB, U.K. (corresponding author e-mail: weisi.guo@cranfield.ac.uk).}
\thanks{We acknowledge funding from EPSRC grant CoTRE - Complexity Twin for Resilient Ecosystems (EP/R041725/1).}}

\maketitle

\begin{abstract}
Resilience characterizes a system's ability to retain its original function when perturbations happen. In the past years our attention mainly focused on  small-scale resilience, yet our understanding of resilience in large-scale network considering interactions between components is limited. Even though, recent research in macro and micro resilience pattern has developed analytical tools to analyze the relationship between topology and dynamics across network scales. The effect of uncertainty in a large-scale networked system is not clear, especially when uncertainties cascade between connected nodes. In order to quantify resilience uncertainty across the network resolutions (macro to micro),an arbitrary polynomial chaos (aPC) expansion method is developed in this paper to estimate the resilience subject to parameter uncertainties with arbitrary distributions. For the first time and of particular importance, is our ability to identify the probability of a node in losing its resilience and how the different model parameters contribute to this risk. We test this using a generic networked bi-stable system and this will aid practitioners to both understand macro-scale behaviour and make micro-scale interventions.
\end{abstract}


\IEEEpeerreviewmaketitle

\section{Introduction}
\IEEEPARstart{I}{n} a connecting world, local dynamics of components in networked systems like critical infrastructure (CI) systems, ecosystems, biological systems etc. affect each other through interactions among components and all together result in a more sophisticated multi-scale network wide dynamics. Example include a water distribution network couples local pumps and reservoirs to deliver supply via Navier-Stokes dynamics \cite{Pagani20}, an electric grid that uses power-flow equations, a fully loaded structure that connects beams and joints via the Ramberg–Osgood equation, a spatially stochastic wireless network that performs traffic load balancing \cite{moutsinas2019probabilistic}, or a fibre optic network that connects optic switches via the Nonlinear Schrodinger's dynamic.

\subsection{Network Resilience Modeling}
Research on resilience of dynamic networks has been widely applied in a wide range of fields from nature to man-made network including blackout in power systems \cite{arghandeh2016definition} to loss of biodiversity in ecology \cite{kaiser2017ecosystem}. Due to the different research context, the concept of resilience is different in different fields \cite{liu2020network}. Up to now, over 70 definitions of resilience have appeared in scientific research \cite{fisher2015more}. In this paper, resilience is defined as the ability of a system to maintain its original functionality when perturbations happen \cite{gao2016universal}. This ability is of great importance in reducing risks and mitigating damages \cite{cohen2000resilience, sole2001complexity}. Whilst the past research on resilience gives us insight into how a few interacting components (small networks) work \cite{sole2001complexity}, the loss of resilience in large-scale networked systems (e.g. $10^5$ nodes) is difficult to predict and analyse analytically. 

These analytical limitations stem from theoretical gaps: most current analytical methods of resilience are appropriate for analyzing models with a high degree of homogeneity which enables mean field to be applied \cite{gao2016universal}. Whilst this has given us insight into the coupling relationship between topology and dynamics, it doesn't enable heterogeneous prediction of node level dynamics. Node level is important to make critical interventions to specific components whilst preserving our multi-scale understanding of general system behaviour. In order to precisely identify the node-level resilience function, a sequential heterogeneous mean field estimation approach is proposed recently \cite{moutsinas2018node}.

\subsection{Uncertainty in Network Resilience}
Mathematical models play an important role in representing physical systems when we analyze the dynamics of networked systems. Actually, a lot of uncertainty exists behind the mathematical models of many practical problems in the real world, which makes these problems more difficult and complex to analyze. These uncertainty may stem directly from incomplete information of the system or measurement noise of initial data as well as from parameters of models whose values are not known exactly \cite{pulch2019uncertainty} \cite{prince2019parametric}. In order to know the effect of arbitrary parameters uncertainty on the network-level resilience, our previous work introduced a polynomial chaos (PC) method \cite{moutsinas2020uncertainty} to understand macro-scale network wide resilience loss uncertainty. However, we still do not know the effect of parameters uncertainty on node-level resilience, especially as parameters uncertainty may cause different effects on nodes in a network. This can paint a different picture to that of the overall macro-scale network behaviour. That is to say, a macro-scale resilient network may hide non-resilient behaviour at the micro-level, which if not addressed in time can cause long term issues.

\subsubsection{Uncertainty Modeling Review}
Since the uncertainty is widespread in practical problems in real world and has an effect on systems' performance, how to quantify these uncertain factors  is the main purpose of research on Uncertainty Quantification (UQ). UQ methods mainly includes: Monte Carlo Methods \cite{fishman2013monte}, Perturbation Methods  \cite{zhao2014quantifying}, Moment Equation Methods \cite{zhang2001stochastic}, Polynomial approximation methods \cite{wiener1938homogeneous}. 

Due to the high accuracy and computational efficiency comparing with traditional Uncertainty Quantification (UQ) methods like Monte Carlo method, Polynomial Chaos Expansion (PCE) method has been widely used in dynamic systems \cite{shen2020polynomial}. For example, a Polynomial Chaos Expansion method was used to estimate the dynamic response bounds of nonlinear system with interval uncertainty \cite{wang2020polynomial}. A Polynomial Chaos Expansion method was applied to analysis the effect of uncertainty in parameters on the received signal concentration in molecular signals \cite{abbaszadeh2018uncertainty}. The PCE was initially proposed to analyze stochastic processes  based on Hermite polynomials, which were only suitable for random variables (r.v.) following Gaussian distribution \cite{wiener1938homogeneous}. However, uncertainty does not always obey the Gaussian distribution. Whilst a normal score transformation could be used to solve this problem \cite{wackernagel2013multivariate}, but can lead to slow convergence \cite{xiu2002wiener}. To solve this problem, the generalized polynomial chaos (gPC) has been developed \cite{xiu2002wiener} \cite{xiu2003modeling}. The gPC extends PCE toward a broader range of applications which could be used encompassing the more general Gamma distribution, Beta distribution, and many other flexible distribution functions. This is further advanced to consider stochastic processes represented by r.v. of any probability distributions \cite{wan2006multi}.

The methods mentioned above need to know the detailed information of the involved probability density functions. While, information about distribution is usually difficult to know or incomplete in practical applications. In different models or circumstances, statistical information of parameters may exist in many different forms. They could be discrete, continuous, or discretized continuous, even exist analytically in the probability density distribution (PDF) or numerically as a histogram. There are two main reasons that limit the widespread use of the above methods. The first reason is that there exist strict restrictions in most cases when they are used. The second one is that the  information of problems to be solved is not always complete and perfect \cite{oladyshkin2012data}.

\subsubsection{Arbitrary Polynomial Chaos}

\subsection{Novelty and Contribution}
While some scientists have begun to research the network resilience, research in estimating node-level resilience considering uncertainty is rarely. In real world, ignoring uncertainty may lead to deviation or even error when estimating resilience of a system as well as a node. Besides, we still don not know the role that uncertainty plays in the nonlinear dynamics system--whether the uncertainty will prevent or accelerate the loss of resilience of system in macro and micro level. Therefore, under this circumstance, it is nature for us to consider uncertainty when estimating resilience of each node in network with nonlinear dynamics.

This paper addresses the lack of uncertainty quantification in the multi-scale resilience of complex networks with nonlinear dynamics. The novelty is to enable parameter uncertainty that follow arbitrary distributions and estimate the resilience of the whole network and each node. To achieve this, a multi-dimensional arbitrary polynomial chaos (aPC) method is developed in this paper to quantify uncertain factors. Besides, we compare dynamics with certain parameters to dynamics with uncertainty when estimating the micro and macro resilience. By analyzing effects of parameters and network topology with uncertainty on the multi-scale resilience of dynamic network, we will better understand large-scale dynamic networks.

\begin{figure*}[ht]
\centering
\resizebox*{14cm}{!}{\includegraphics{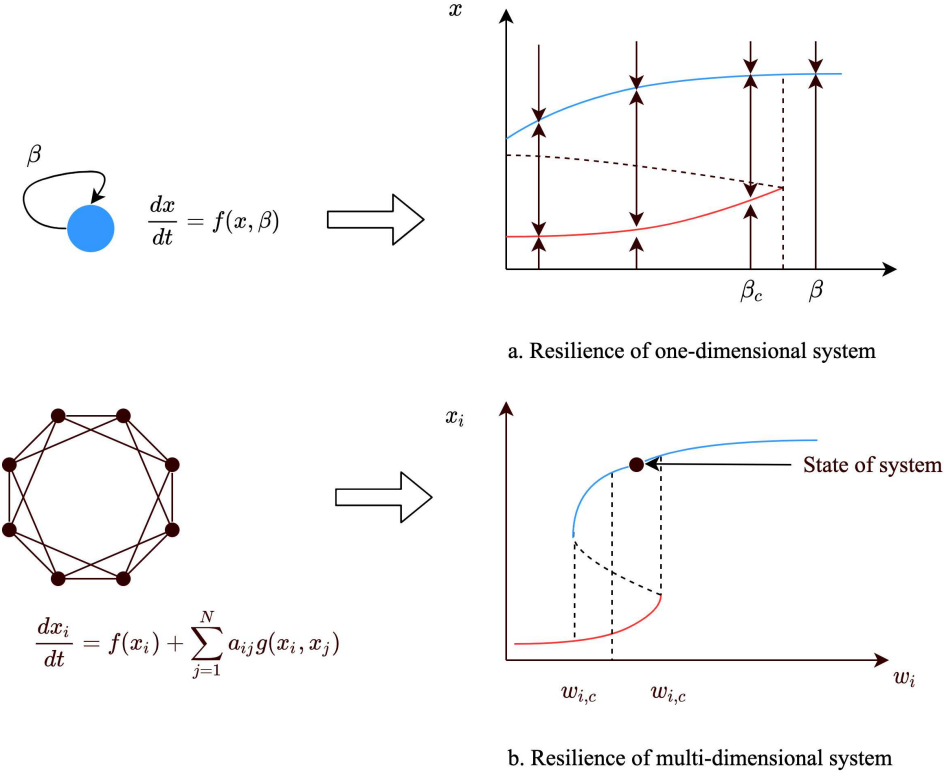}}
\caption{It shows the dynamics of one-dimensional system and multi-dimensional system. In a one-dimensional system,
resilience is characterised  by the function $x(\beta)$. When $\beta > \beta_{c}$, only one stable fixed point (blue) exist in the system. When $\beta < \beta_{c}$, two (or more) stable fixed points will appear, which maps to a desirable state (blue) and an undesirable state (red). (b) In a multi-dimensional dynamic system, dynamics of system is captured by the complex weighted network $w_i$ instead of $\beta$. $w_i$ is affected by both environmental conditions and the interaction strength. }
\label{fig:1}
\end{figure*}

\section{System Setup}

\subsection{Node Level Nonlinear Dynamics and Resilience}
\begin{figure}[ht]
\centering
\resizebox*{8cm}{!}{\includegraphics{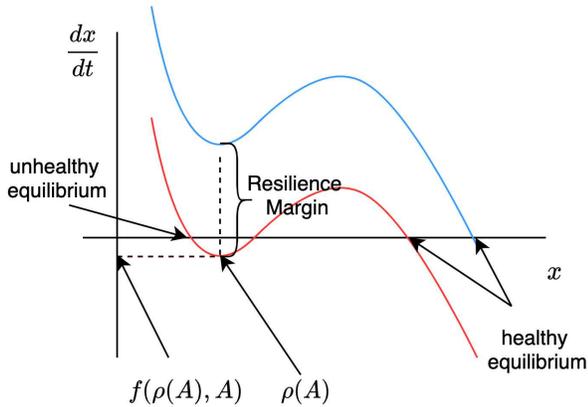}}
\caption{Red line describes a system with more than one stable equilibrium (healthy one and unhealthy one both exist). Blue line describes a system with only one stable equilibrium.}
\label{fig2}
\end{figure}

The resilience behavior of a one-dimensional nonlinear dynamic system in ecology \cite{may1977thresholds}, engineering \cite{lyapunov1992general} etc. could be characterized by the equation:
\begin{equation}\label{equ1}
\dot{x}=f(\beta,x)
\end{equation}
where \(f(\beta,x)\) shows the system's dynamics, and the parameter \(\beta\) shows environment conditions (show in Figure (\ref{fig:1})). One of the stable fixed points, \(x_0\) of equation (\ref{equ1}) could be found by
\begin{equation}\label{equ2}
    f(\beta, x_0)=0
    \end{equation}
    \begin{equation}\label{equ3}
        \left. \frac{df}{dx} \right|_{x=x_0}<0
\end{equation} where $f$ is smooth and equation (\ref{equ2}) provides the system's steady state and equation (\ref{equ3}) guarantees its linear stability. We assume that a stable equilibrium $x_e>0$ always exists which is away from the origin. Besides, the bifurcation may occur near to the origin shown in Figure (\ref{fig2}). We define two different kinds of stable equilibrium: \textit{healthy} equilibrium and \textit{unhealthy} equilibrium. The \textit{healthy} equilibrium is far from the origin and it is a desired state of the system. The \textit{unhealthy} equilibrium is near to the origin and it is an undesired one. \textbf{Resilience} in this general case is defined by a healthy and an unhealthy equilibrium.  Only when does the healthy equilibrium exist in the system, the system is resilient. While, if in the system healthy and unhealthy equilibrium exist at the same time, the system will transit from the desired stable fixed point to the undesired one, which indicates the loss of resilience in the system.

\begin{figure*}[ht]
\centering
\resizebox*{14cm}{!}{\includegraphics{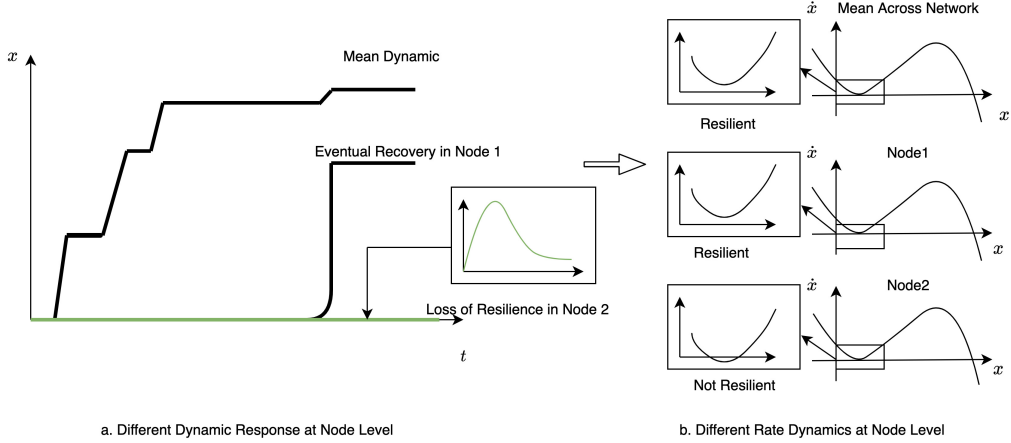}}
\caption{Similar Network Dynamics Hide Different Node Dynamics. It shows different dynamic response at node level. While the mean dynamic shows the network is resilient, node 1 and node 2 have different dynamics.  (a) Node 1 recoveries its resilience eventually, but node 2 loses its resilience. (b)Node 1 only has one healthy equilibrium, but node 2 has a healthy equilibrium and an unhealthy equilibrium. }
\label{fig3}
\end{figure*}

\begin{figure*}[ht]
\centering
\subfloat[Resilience Bounds and Uncertainty Region with certain parameters]{%
\resizebox*{14cm}{!}{\includegraphics{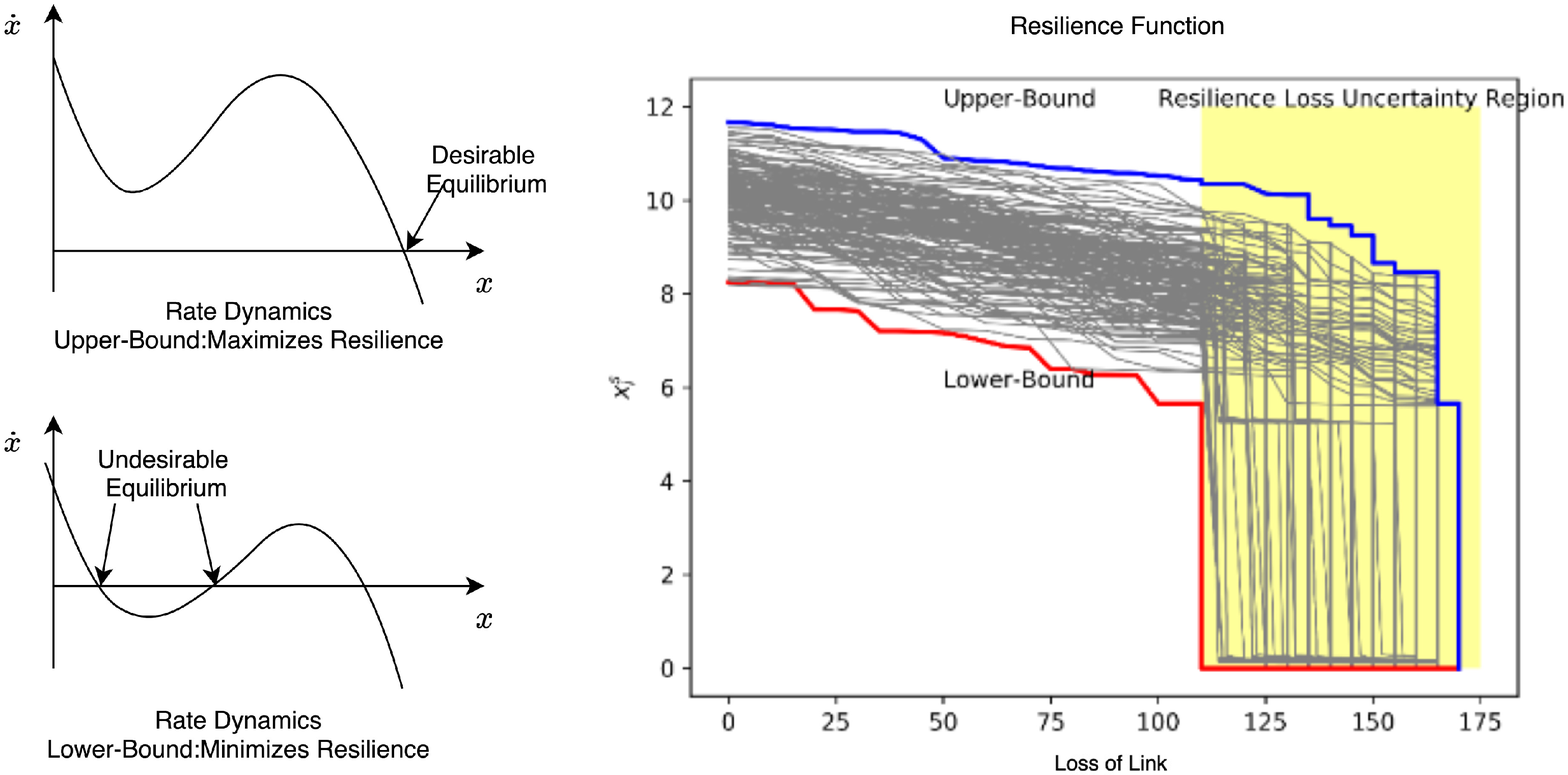}}}\hspace{5pt}
\subfloat[Critical Resilience Value at Node Level with certainty parameters]{%
\resizebox*{14cm}{!}{\includegraphics{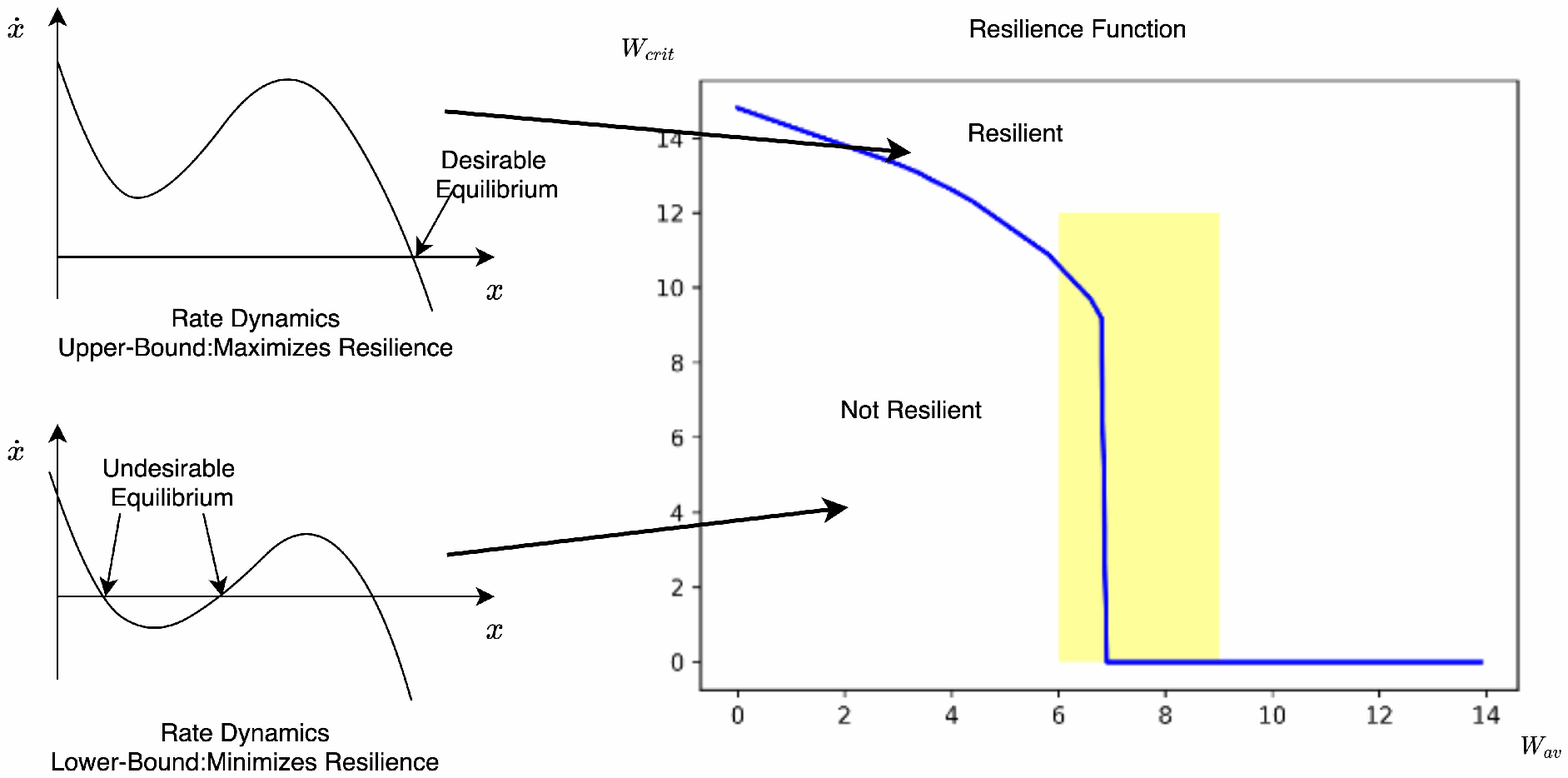}}}
\caption{Critical Resilience Value Identifies Vulnerable Nodes with Certain Parameters. (a) Resilience Bounds shows the Upper-Bound and Lower-Bound of equilibrium when links removed. In this figure, it explicitly predicts when the loss of resilience will happen. (b) Critical Resilience shows the relationship between average weight value of network and critical weight value. When $w_i>w_\text{crit}$, the node is resilient, otherwise it is not.} \label{fig4}
\end{figure*}

\begin{figure*}[ht]
\centering
\subfloat[The effect of uncertain parameters on network when average weight is 7.]{%
\resizebox*{8cm}{!}{\includegraphics{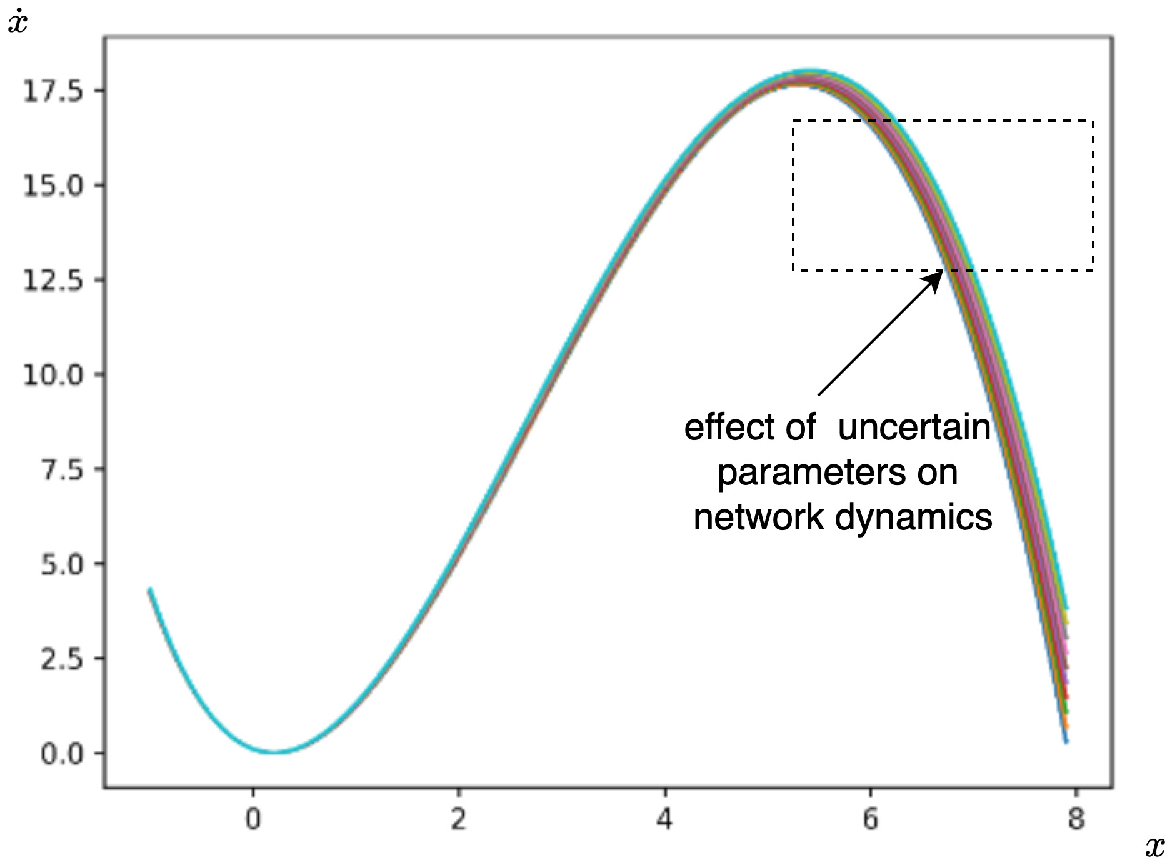}}}\hspace{5pt}
\subfloat[The effect of uncertain parameters on a node's equilibrium]{%
\resizebox*{8cm}{!}{\includegraphics{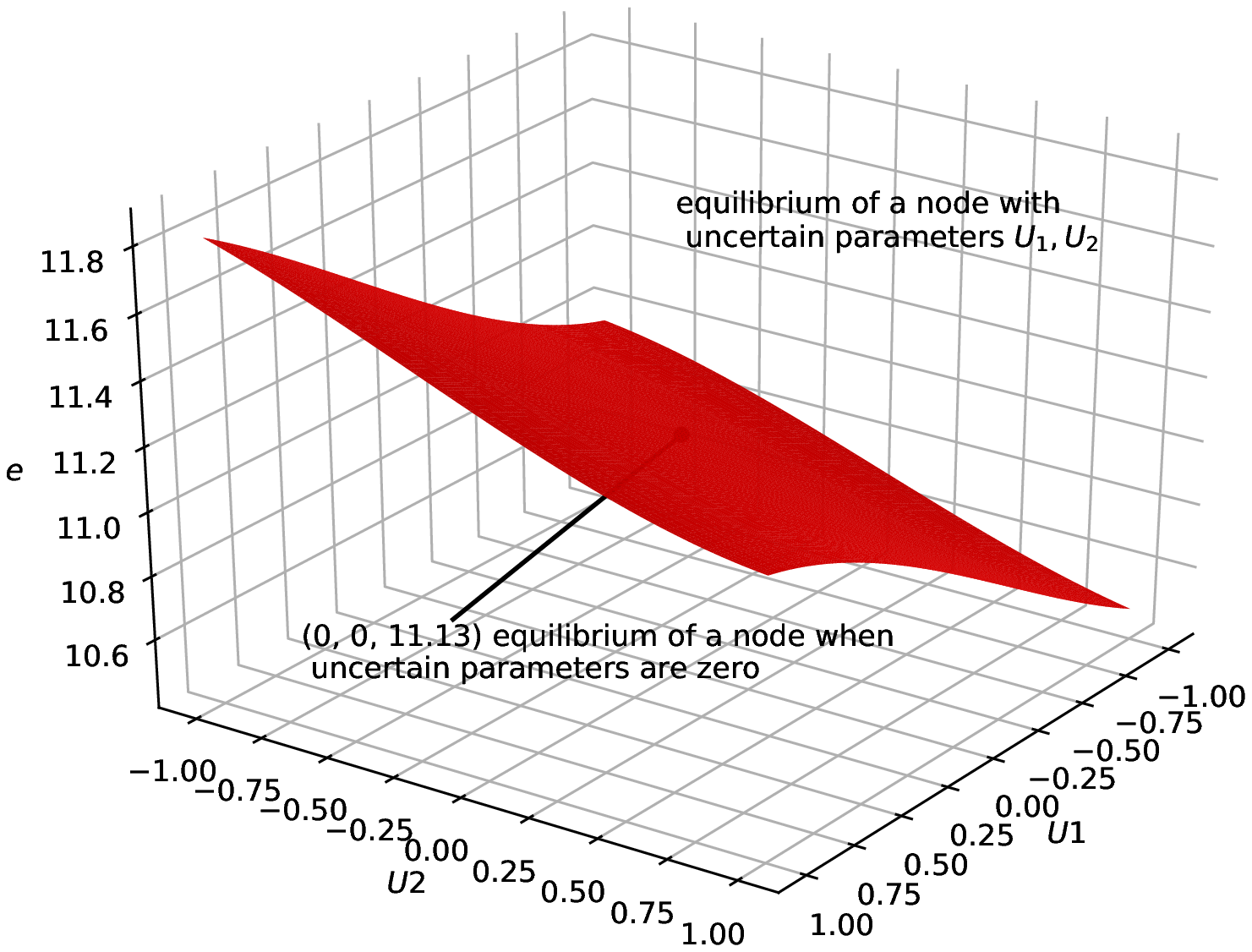}}}
\caption{The effect of uncertain parameters on network and node. In system with certain parameters, network and node-level dynamics are certain. While in a system with uncertain parameters, network and node-level dynamics are uncertain. In (b) $e$ represents the equilibrium of a node, $U_1, U_2$ are uncertain parameters.} \label{fig5}
\end{figure*}

\subsection{Network Level}
Networked system often consists of a large number of components interacting with each other through the network(show in Figure (\ref{fig:1}) (b)). 
\begin{equation}\label{equ4}
    \dot{x_i} = f(x_i, a_i)+\sum_{j=1}^n a_{ji} g(x_i,x_j, b_{ij})
\end{equation} 
Node $i$'s behavior is characterised by a self-dynamic $f(\cdot)$ and a coupling dynamic $g(\cdot)$. $\textbf{A}$ and $\textbf{B}$ both are vectors of parameters. $\textbf{A} = \left\{a_1, ..., a_i\right\}$, $\textbf{B} = \left\{b_{11}, ..., b_{ij}\right\}$. So we can rewrite equation (\ref{equ4}):
\begin{equation}\label{equ5}
    \dot{X}=F(X, \textbf{A}, \textbf{B}),
\end{equation} where $F:\textbf{R}^N\xrightarrow{}\textbf{R}^N$ defined by equation (\ref{equ4}). Let $w_i$ be the weighted in-degree of vertex $v_i$, i.e. 
\begin{equation}\label{equ6}
    w_i = \sum_{j=1}^N a_{ji},
\end{equation} and $w_\text{av}$ represents the average of all weighted in-degrees. The weighted out-degree of vertex $v_j$ is represented by $w_j^\text{out}$ , i.e.
\begin{equation}\label{equ7}
    w_j^\text{out} = \sum_{i=1}^N a_{ji}.
\end{equation}

Similarly, let $d_i$ be the in-degree of $v_i$ and  $d_i^\text{out}$ be its out-degree. Generally, the relationship between topology (e.g. properties of $\textbf{M}_{ij}$) and resilience of network is still not very clear. One way to solve this problem is to compress the multi-dimensional dynamics to  one-dimensional dynamics and map the overall effective dynamics of a networked system to its topology. Indeed,a  common network-level effective dynamics may hide different node-level dynamics of different nodes (shown in Figure (\ref{fig3})). In order to understand node-level resilience and dynamics, a sequential estimation approach is proposed to solve this problem  \cite{moutsinas2018node}. However, we still do not know the effect of uncertainty parameters on the resilience at node-level.

\section{Approach and Methodology}

To answer this question, an arbitrary polynomial chaos expansion method is proposed to estimate the resilience at node-level with uncertainty. We do so by defining arbitrary uncertainty distributions on the network dynamic parameters.

\subsection{Dynamic network with uncertainty}
Uncertainty may exit in self-dynamics, coupling dynamic or topology of networked dynamic system. It is assumed that parameters in dynamics function could be represented by random variables. What's more, parameters' value are assumed to be within a range of their real values. Therefore, we get $a_i= a_i(1+e_1u_i)$, $b_{ij}=b_{ij}(1+e_2v_{ij})$, $\textbf{M}=\textbf{M}(1+e_3r)$, where $u_i, v_{ij}, r$ are r.v. uniform in $[a, b]$ and $e_1, e_2, e_3$ are constants. $\textbf{U} = \left\{u_1, ..., u_i\right\}$. $\textbf{V} = \left\{ v_{11}, ..., v_{ij} \right\}$. The dynamic network' mathematics model with uncertainty can be written as:
\begin{equation}\label{equ8}\begin{split}
    &\dot{x_i}=f(x_i,a_i(1+e_1u_i))+\\ &\sum_j^na_{ji}(1+e_3r)g(x_i,x_j, b_{ij}(1+e_2v_{ij}))
    \end{split}
\end{equation}

\subsection{Sequential Heterogeneous Mean Field Estimation}
A mean field estimation is used to drive sequential substitution and estimation of node-level resilience.

\subsubsection*{\textbf{Step 0}}
\label{step0} 
Firstly, we could estimate the equilibrium $e^{\left\{0\right\}}$ of the mean field dynamics of the networked system through a homogeneous average degree $w_\text{av}=\frac{1}{N}\sum_{i=1}^N\sum_{j=1}^Na_{ij}$ or a weighted average degree \cite{gao2016universal}
\begin{equation}\label{equ9}
    w_\text{av} = \frac{<w^\text{out}w^\text{in}>}{<w^\text{out}>}
\end{equation}
$w^\text{out}=(w_1^\text{out}, w_2^\text{out}, ..., w_N^\text{out})$ is a vector of  weighted out-degrees, $w^\text{in}=(w_1^\text{in}, w_2^\text{in}, ..., w_N^\text{in})$ is a vector of weighted in-degrees, and $<w^\text{out}>=(\frac{1}{N}\sum_{i=1}^Nw_i^\text{out})$ is the average weighted out-degree. We define $\textbf{1}:={1,...1}\in R^N$. The mean field approximation of the equilibrium can be calculated according to equation (\ref{equ10}).
\begin{equation}\label{equ10}\begin{split}
    &\Xi :=Mean[F(x\textbf{1, A, B})] =\\ &\frac{1}{N}\sum_{i=1}^N(f(x,a_i))+\frac{1}{N}\sum_{i=1}^Nw_\text{av} g(x,x, b_{ij})
    \end{split}
\end{equation}

$\Xi(x)$ depends on \textbf{A} and \textbf{B}. \textbf{A} and \textbf{B} are both vectors of r.v., for any $x$. Therefore, $\Xi(x)$ is a function depending on the random variable $x$. And we find $x$ which satisfies the function $\Xi(x)=0$.

For fixed $x$, $f(x, a_i)$ is a function depending on iid r.v. $a_i$. We set
\begin{equation}\label{equ11}
    \mu_{f(x)}:=\textbf{E}[f(x, a_i)]
\end{equation}
\begin{equation}\label{equ12}
    \delta_{f(x)}:=\sqrt{\textbf{Var}[f(x, a_i)]}
\end{equation}
According to Central Limit Theorem (CLT), for big enough $n$, $\frac{1}{n}\sum_{i=1}^nf(x, a_i)$ can be approximated by a normally distributed random variable with mean $\mu_{f(x)}$ and standard deviation $\frac{1}{n}\delta_{f(x)}$, i.e
\begin{equation}
    \frac{1}{n}\sum_{i=1}^nf(x, a_i)\sim\textbf{N}(\mu_{f(x)}, \frac{1}{n}\delta_{f(x)}^2)
\end{equation}\label{equ13}
$g(x,x,b_{ij})$ is a function depending on random variable $x$. We set
\begin{equation}\label{equ14}
    \mu_{g(x)} := \textbf{E}[g(x,x,b_{ij})]
\end{equation}
\begin{equation}\label{equ15}
    \delta_{g(x)}:=\sqrt{\textbf{Var}[g(x,x,b_{ij})]}
\end{equation}
Then we can get
\begin{equation}\label{equ16}
    \frac{1}{n}\sum_{i,j=1}^nM_{ji}g(x,x,b_{ij})\sim\textbf{N}(\frac{m}{n}\mu_{g(x)},\frac{m}{n^2}\delta_{g(x)}^2)
\end{equation}

From the mentioned above, we know that $\Xi(x)$ is the sum of 2 normally distributed r.v.. Then we can get 
\begin{equation}\label{equ17}
    \Xi(x)\sim\textbf{N}(\mu_{f(x)}+\frac{m}{n}\mu_{g(x)},\frac{1}{n}\delta_{f(x)}^2+\frac{m}{n^2}\delta_{g(x)}^2)
\end{equation}
$\Xi_\alpha(x)$ could be realised by drawing $\zeta_\alpha$ from $\textbf{N}(0,1)$ and we can get
\begin{equation}\label{equ18}
    \Xi_\alpha(x) = \mu_{f(x)}+\frac{m}{n}\mu_{g(x)}+\sqrt{\frac{1}{n}\delta_{f(x)}^2+\frac{m}{n^2}\delta_{g(x)}^2}\zeta_\alpha
\end{equation}
It is assumed that every realisation of $\Xi(x)$ has the shape described in Figure (\ref{fig2}). The equilibrium $e^{\left\{0\right\}}$  could be calculated from equation (\ref{equ18}). Since $\zeta_\alpha$ is a random variable which is normally distributed, a polynomial chaos expansion (PCE) could be used to approximate the equilibrium $e^{\left\{0\right\}}$. $\textbf{P}(e^{\left\{0\right\}})$. We calculate the smallest positive root $\rho^{\left\{0\right\}}$ of $\Xi^{'}(x)$ and set $\tau^{\left\{0\right\}}=\Xi(\rho^{\left\{0\right\}})$.

Since $\Xi(x)$ is a random variable, $\rho^{\left\{0\right\}}$ and $\tau^{\left\{0\right\}}$ are functions based on this random variable. Meanwhile, $\tau^{\left\{0\right\}}$ is an indicator for the saddle-node bifurcation, which means that whether the system is resilient could be judged by $\tau^{\left\{0\right\}}$. To be specific, for a given realization of $\zeta_\alpha$, if $\tau_\alpha^0>0$, then only one equilibrium (the healthy one) exists in the system and the system is resilient. Otherwise, if $\tau_\alpha^0<0$, three equilibrium exist in the system including healthy and unhealthy equilibrium. Then the dynamics is non-resilient. $\textbf{P}(\tau>0)$ represents the probability of the system being resilient. We use PCE to estimate $\tau(\zeta)$. This PCE is represented by $\widetilde{\tau}_n(\zeta)$ and we set 
\begin{equation}\label{equ19}
    pos(x)= \left\{
             \begin{array}{lr}
             1  \quad \textup{if} \quad x>0 &  \\
             0  \quad \textup{otherwise} &  
             \end{array}
\right.
\end{equation}
Then, the probability of resilience could be calculated 
\begin{equation}\label{equ20}
    \frac{1}{\sqrt{2\pi}} \iint\limits_{-\infty}^{+\infty}pos(\widetilde{\tau}_n(\zeta))\,d\zeta
\end{equation}

\subsubsection*{{\textbf{Step 1}}}
\label{step1}
The mean field approximation is used as an initial guess to estimate the dynamics of each node:
\begin{equation}\label{equ21}
   \dot{x_i} = f(x_i,a_i) + w_ig(x_i,e^{\left\{0\right\}},b_{ij}) = 0
\end{equation}
The solution of equation (\ref{equ21}) is a function of $w_i$, i.e. $\chi^{\left\{1\right\}}(w_i)$. Then we have first order approximation  $e^{\left\{1\right\}}_i=\chi^{\left\{1\right\}}(w_i)$. Since parameters $a_i, b_{ij}$ does not always belong to common distribution like Gaussian distribution, Binomial distribution etc. We need to use the arbitrary polynomial chaos (aPC) \cite{oladyshkin2012data} to approximate $e^{\left\{1\right\}}_i$ and its distribution. 

\subsubsection*{\textbf{Step 2}}
\label{step2} 
The previous approximation could be used to estimate the effect of the graph on a vertex. The effect of an in-edge on the dynamics of vertex $i$ is $g(x_i,x_j)$ and the probability of of a vertex $j$ is on the other side of the in-edge is proportional to its out-degree. So the average effect is $\sum_{j=1}^Nd_j^\text{out}g(x_i,x_j,b_{ij})/\sum_{j=1}^Nd_j^\text{out}$. To approximate mean effect of the neighbours, components in $g(\cdot)$ are weighted by $d^\text{out}$. Therefore, the previous step's estimation could be used to make further estimation. And we can approximate the equilibrium of each node from 
\begin{equation}\label{equ22}
    \dot{x}=f(x_i)+w_i\frac{\sum_{j=1}^Nd_j^\text{out}g(x_i,e_j^{\left\{1\right\}})}{\sum_{j=1}^Nd_j^\text{out}}=0
\end{equation}
Notice that the solution depends on $w_i$. So the second order approximation is $e^{\left\{2\right\}}_i=\chi^{\left\{2\right\}}(w_i)$. Also, $e^{\left\{2\right\}}$ and its distribution could be approximated by aPC. 

\subsubsection*{\textbf{Step 3 to n}}
\label{step3} We repeat the above steps with each approximation calculated in the previous step.

\subsection{Arbitrary Polynomial Chaos Expansion}

\subsubsection{One-Dimensional aPC}
$\Xi$ is defined as a random variable with PDF $w$. Let us consider a stochastic model $X=\phi(\Xi)$. $\phi$ is a function that is square integrable on $\textbf{R}$ with a weight function $w$. For a stochastic analysis of $X$, the model $\phi(\Xi)$ may be expanded as follows:
\begin{equation}\label{equ23}
    \phi(\Xi)=\sum_{i=0}^dc_iP^{(i)}(\Xi)
\end{equation} 
$c_i$ are expansion coefficients and $P^{(i)}(\Xi)$ are the polynomials which forms the basis $\left\{P^{(0)}, P^{(1)}, P^{(2)}, ..., P^{(i)}\right\}$. $P^{(i)}(\Xi)$ is orthogonal with respect to $w$. In aPC, $w$ may have an arbitrary form, which is different general PCE methods. The basis $\left\{P^{(0)}, P^{(1)}, P^{(2)}, ..., P^{(i)}\right\}$ need to be formed specifically according to the statistics information of $w$.

\subsubsection{Multi-Dimensional aPC}
In some cases, the number of input parameters is more than one, i.e. $\Xi=\left\{\Xi_1, \Xi_2,...,\Xi_N\right\}$. The model output $X$ could be approximated by a multivariate polynomial expansion:
\begin{equation}\label{equ24}
    \phi(\Xi_1, \Xi_2, ..., \Xi_N)=\sum_{i=1}^Mc_i\Phi_i(\Xi_1, \Xi_2, ..., \Xi_N).
\end{equation}
In equation (\ref{equ24}), the number of input parameters is $N$ and the number of $M$ is decided by the formula $M = (N+d)!/(N!d!)$, where $d$ represents the order of expansion. Here, we need to construct the orthogonal polynomial basis $\Phi_i$ for $\Xi_1, \Xi_2,...,\Xi_N$. Assuming that the input parameters are independent, the multi-dimensional basis can be constructed as a simple product of the corresponding univariate polynomials
\begin{equation}\label{equ25}\begin{split}
    &\Phi_i(\Xi_1, \Xi_2, ..., \Xi_N)=\prod_{j=1}^NP_j^{(\alpha_j^i)}(\Xi_1, \Xi_2,...,\Xi_N), \\
    &\sum_{j=1}^N\alpha_j^i\leq M, \quad i=1,...,N,
\end{split}
\end{equation} 
In equation (\ref{equ25}), $\alpha_j^i$ is a multivariate indicator with the information on how to list all possible products of individual univariate basis functions. 
$\alpha_j^i$ is a multivariate index containing information about how to enumerate all possible products of individual univariate basis functions. Here, we give an example to illustrate it. For example, we assume that $d=2, N=2$, then $M=6$. Therefore, according to equation (\ref{equ25}), the basis functions are $\left\{1, \Xi_1, \Xi_2, \Xi_1\Xi_2, \Xi_1^2, \Xi_2^2\right\}$.

We define the polynomial $P^{(k)}(\Xi)$ of degree $k$ in the random variable $\Xi$:
\begin{equation}\label{equ26}
    P^{(k)}(\Xi)=\sum_{i=0}^kP_i^{(k)}\Xi^i, k \in \left[0, d\right]
\end{equation}
where $P_i^{(k)}$ are coefficients in $P^{(k)}(\Xi)$.

The key of the aPC method is to construct the polynomials in equation (\ref{equ26}) to form an orthonormal basis for arbitrary distributions which could be discrete, continuous, raw data sets or by their moments. we define the orthonormality for polynomials $P^{(k)}$ and $P^{(l)}$ as 
\begin{equation}\label{equ27}
    \int P^{(k)}(\Xi)P^{(l)}(\Xi)dw(\Xi)= \left\{
             \begin{array}{lr}
            0 \quad  \forall k\neq l &  \\
            1 \quad else &  
             \end{array}
\right.\\
\end{equation}
Here we assume that the leading coefficients of all polynomials: $P_k^{(k)}=1 \quad \forall k$.
The $k$th raw (crude) moment of the random variable $\Xi$ is defined as 
\begin{equation}\label{equ28}
    \mu_k = \int\Xi^kdw(\Xi)
\end{equation}

The relationship between raw moments of $\Xi$ and their coefficients can be written in matrix form (the detail process could be seen in \cite{oladyshkin2012data}):
\begin{equation}\label{equ29}
    \left[ \begin{array}{cccc}
\mu_0 & \mu_1 &... & \mu_k\\
\mu_1 & \mu_2 &... & \mu_{k+1}\\
\vdots &\vdots &\vdots & \vdots\\
\mu_{k-1} & \mu_k &... & \mu_{2k-1}\\
0 & 0 & ... & 1
\end{array} 
\right ]
\left[ \begin{array}{c}
P_0^{(k)} \\
P_1^{(k)} \\
\vdots\\
P_{k-1}^{(k)}\\
P_k^{(k)}
\end{array} 
\right ]
=
\left[ \begin{array}{c}
0 \\
0 \\
\vdots\\
0\\
1
\end{array} 
\right ]
\end{equation}
For multi-dimensional r.v., the polynomial $P_j^{(k)}(\Xi_j)$ is defined as:

\begin{equation}\label{equ30}
    P_j^{(k)}(\Xi_j)=\sum_{i=0}^kP_{i,j}^{(k)}\Xi_{j}^i  
\end{equation} 
and the unknown polynomial coefficients $P_{i,j}^{(k)}$ can be defined from the following matrix equation \cite{oladyshkin2011concept}:
\begin{equation}\label{equ31}
    \left[ \begin{array}{cccc}
\mu_{0, j} & \mu_{1, j} &... & \mu_{k,j}\\
\mu_{1, j} & \mu_{2, j} &... & \mu_{k+1, j}\\
\vdots &\vdots &\vdots & \vdots\\
\mu_{k-1, j} & \mu_{k, j} &... & \mu_{2k-1, j}\\
0 & 0 & ... & 1
\end{array} 
\right ]
\left[ \begin{array}{c}
P_{0,j}^{(k)} \\
P_{1,j}^{(k)} \\
\vdots\\
P_{k-1,j}^{(k)}\\
P_{k, j}^{(k)}
\end{array} 
\right ]
=
\left[ \begin{array}{c}
0 \\
0 \\
\vdots\\
0\\
1
\end{array} 
\right ]
\end{equation}
We now show the results of a real system case study to illustrate how the aPC framework can be used.

\begin{figure*}[ht]
\centering
\resizebox*{14cm}{!}{\includegraphics{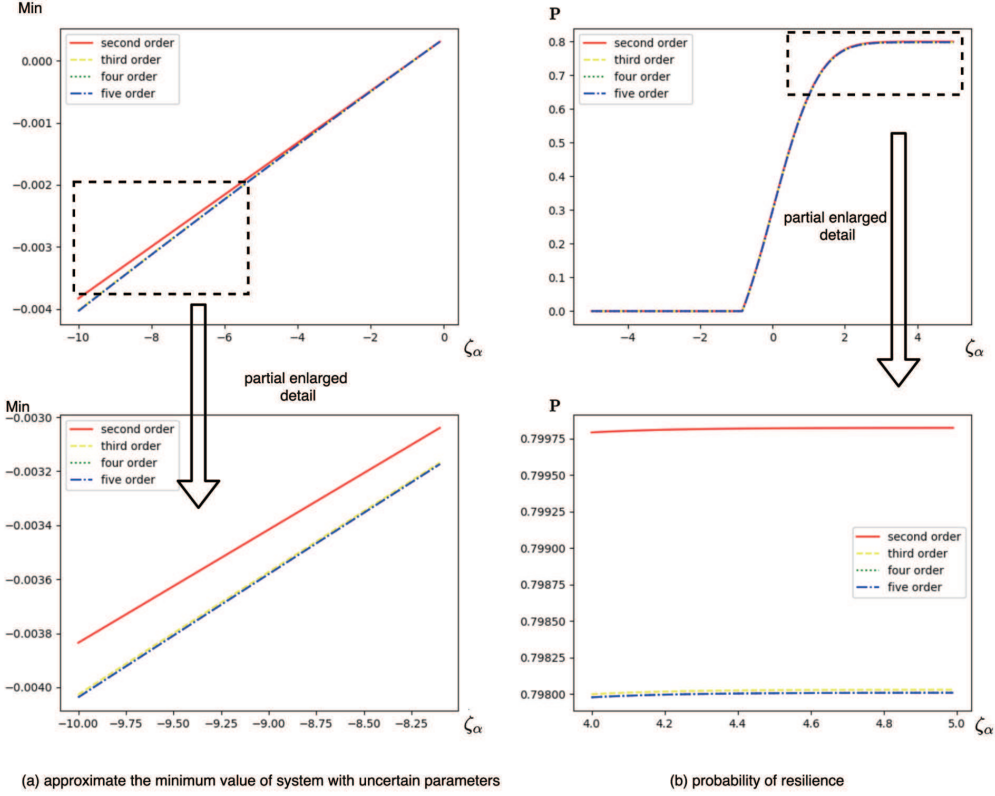}}
\caption{Approximate resilience of system by Polynomial Chaos Expansion. We truncate the series to arbitrary order $N$ from 2 to 5. (a) approximate the minimum value of system by PCE. (b) approximate the probability of resilience. It is clear that there is a significant difference in results between $N=2$ and $N=3,4,5$ in (a) and (b).}
\label{fig6}
\end{figure*}

\section{Results for Bi-Stable Systems}

Bi-stable dynamical systems are common across social (e.g. population logistic model \cite{Lundqvist10}), ecological (e.g. soil health \cite{Corstanje18}), climate (e.g. ocean circulation \cite{Lenton04}), and human conflict systems \cite{Aquino19}. There exists a stable undesirable state (e.g. population collapse or conflict) and a stable desirable state (e.g. healthy population with collaboration \cite{Ron18}), with an unstable transition brink in between, and this is ideal for demonstrating the concept of resilience and uncertainty. Networks that connect such systems represent a wider interacting ecosystem and often a mutualistic coupling represents positive reinforcing interactions. Interaction examples include gravity, radiation, or Boltzmann Lotka Volterra (BLV) models \cite{Wilson08} frequently use a $x_i \times x_j$ mutualistic attractor component. 

\subsection{Case Study: Ecological Network}

A case of pollinator networks is used to illustrate the dynamics of networked system at micro level and macro level\cite{holland2002population}. $x_i$ represents the abundance of species $i$, which is given by:
\begin{equation}\label{equ32}
    \frac{dx_i}{dt}=B_i+x_i(1-\frac{x_i}{K_i})(\frac{x_i}{C_i}-1)+\sum_j^N a_{ji}\frac{x_ix_j}{D_i+E_ix_i+H_jx_j}
\end{equation}
$B_i$ represents the incoming migration rate of species $i$ from other ecosystems. The second term on right hand shows logistic growth with carrying capacity $K_i$ of the system, and the Allee effect (low abundance ($x_i<C_i$) causes negative growth) \cite{allee1949principles}. The third term is a coupling function which saturates for large $x_i$ or $x_j$ ($j$'s positive contribution to $x_i$ is bounded).

For simplicity, we use homogeneous parameters: $B=0.1, C=1, K=5, D=5, E=0.9, H=0.1$. Besides, it is assumed that some parameters' value has to be within $10\%$ of its mean. Here, we 
set $C = \textbf{E}[C](1+0.1U_1), E=\textbf{E}[E](1+0.1U_2)$, where $U_1, U_2$ are random variables uniform in $[-1, 1]$ ($U_1, U_2$ could be r.v. that follow arbitrary distributions. We only need to know the statistics information of their raw moments.).  The definition of system resilience in this model is the ability of the system to recover species abundance from extinction \cite{moutsinas2018node}. To achieve this, the system should be in the regime where only one equilibrium exists. This because if over one equilibrium exist in the system, the system will be trapped in the state with low abundance, which means that the system can not recover its species abundance and loses its resilience.

In Figure (\ref{fig4}), we show what happened when a network becomes less connected by removing edges. In this case, parameters are certain and the figure explicitly shows the bounds of equilibrium under different perturbation and the regime where loss of resilience happens. Critical function describes resilience regimes which maps macro (network-level) properties (average weighted degree $w_\text{av}$ to micro (local-level) properties (critical resilience value $w_\text{crit}$)). For each $w_\text{av}$, corresponding $w_\text{crit}$ could be calculated from equation (\ref{equ33}). The critical weight, $w_\text{crit}$, is a function of $w_\text{av}$ since it is a function of $e^{\left\{0\right\}}$ and $e^{\left\{0\right\}}$ is a function of $w_\text{av}$. In Figure (\ref{fig4}) (b), we see the graph of $w_\text{crit}$ versus $w_\text{av}$. Since $e^{\left\{0\right\}}$ is discontinuous, $w_\text{crit}$ is also discontinuous.

\begin{equation}\label{equ33}
\dot{x_i} = f(x_i)+w_ig(x_i, e^{\left\{0\right\}}(w_\text{av}))
\end{equation}

In this case, a critical average weight $w_{*}$ is about 7 where bifurcation will happen. When average weight is greater than 7, the system is resilient and almost every node in this system is resilient. The critical weight can reveal some basic properties for the dynamics on a nodal level. For example we see in Figure (\ref{fig4}) (b) that when when $w_\text{av}>w_{*}$, $w_\text{crit}$ is almost 0. This implies that if the system on average is in the resilient region, a vertex will also be in the resilient region even if it is very weakly connected to the rest of the network. However, in the case with uncertain parameters, even if the average weight is greater than 7, the system is possibly not resilient. 
We use aPC to analysis what will happen in the regime where loss of resilience may exist. In Figure (\ref{fig5}), it shows dynamics of the system with uncertain parameters when average weight is 7. We use aPC to approximate the minimum value of the function and whether the system is resilient.

\begin{figure*}[htbp]
\centering
\subfloat[Approximate equilibrium of a node by aPC when we truncate the series to arbitrary orders $N$ from 2 to 5]{%
\resizebox*{8cm}{!}{\includegraphics{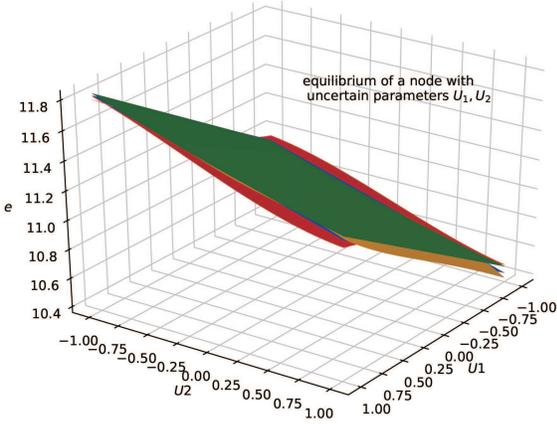}}}\hspace{5pt}
\subfloat[The CDF of equilibrium when  we truncate the series to arbitrary orders $N$ from 2 to 5 ]{%
\resizebox*{8cm}{!}{\includegraphics{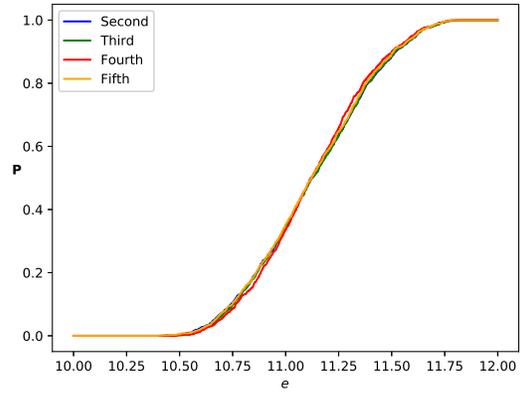}}}
\caption{Approximate equilibrium of a node in the networked system by aPC. In (a), the four color surface , blue, green, red, yellow surface, present different truncation from 2 to 5. $e$ represents the equilibrium of a node, $U_1, U_2$ are uncertain parameters. It can be seen that these surface almost overlap which means that their accuracy are similar. In (b), it shows the CFD when we truncate the series from 2 to 5.} \label{fig7}
\end{figure*}

\begin{figure*}[htbp]
\centering
\subfloat[Probability of resilience when average weight of network is different]{%
\resizebox*{8cm}{!}{\includegraphics{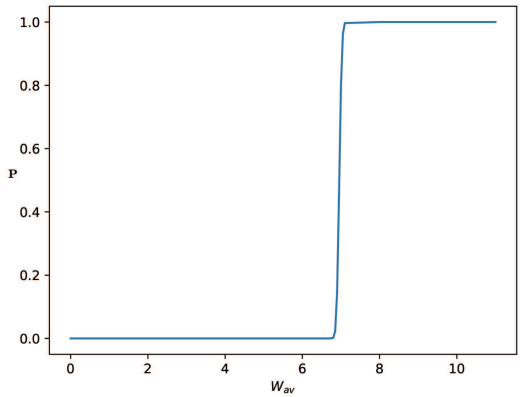}}}\hspace{8pt}
\subfloat[Critical weight of node with different average weight of network]{%
\resizebox*{8cm}{!}{\includegraphics{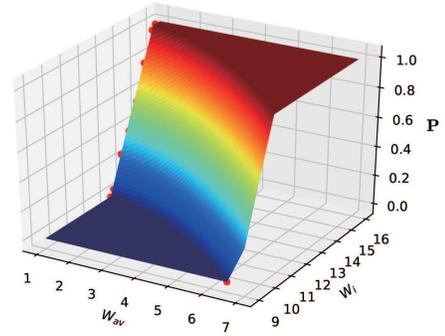}}}

\subfloat[Difference between certain and uncertain parameters when average weight of network changes]{%
\resizebox*{8cm}{!}{\includegraphics{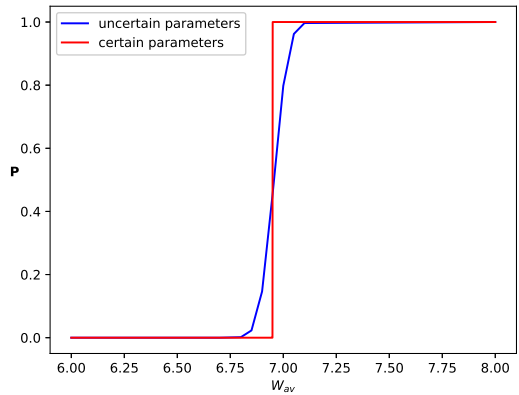}}}\hspace{8pt}
\subfloat[Relationship between critical weight and average weight with certain and uncertain parameters ]{%
\resizebox*{8cm}{!}{\includegraphics{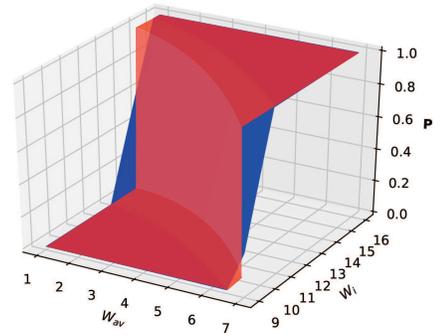}}}
\caption{It shows the effect of uncertainty parameters on resilience of network and each node. (a) (b) show the probability of resilience at network-level and node-level in system with uncertainty. (c) (d) show the difference between resilience with certain parameters and uncertain parameters at network-level and node-level (Blue represents system with uncertain parameters and red represents system with certain parameters).} \label{fig8}
\end{figure*}

\subsection{Analysis on the Effect of Uncertainty}
Firstly, we use the method described in \nameref{step0} to analyse the probability of system to be resilient and approximate the equilibrium. We truncate the series to arbitrary orders $N$ from 2 to 5 (Figure (\ref{fig6})). It is clear that the convergence of the function can be improved with the increase of the polynomial order ($N$). However, with the increasing of the order, much more simulation is needed. Therefore, we have to make a compromise between accuracy and computational efficience.

In Figure (\ref{fig6}), it clearly shows the difference among different orders especially $N=2$. To calculate the probability of resilience, a graph of Cumulative Distribution Function (CDF) with different truncation is shown in Figure (\ref{fig6}) (b). In Figure (\ref{fig6}), when $N=3, 4, 5$, the results are almost the same. However, there is an obvious difference for $N=2$.

Considering the accuracy and computational efficiency, we choose $N=3$ for the polynomial order.  So, we can see the effect of uncertain parameters on system resilience as well as node-level resilience. When parameters are certain and average weight is 7, the system is resilient and nodes are resilient. However, when parameters are uncertain in this case, the probability of resilience of the system is about 0.798. So according to analysis above, some nodes will also possibly lose resilience. 

Second, we use the method in \nameref{step1} and \nameref{step2} to estimate the  equilibrium of each node. The method aPC mentioned above is used to estimate a node's equilibrium and we truncate the series to arbitrary orders $N$ from 2 to 5 shown in Figure (\ref{fig7}). In Figure (\ref{fig7}) (a), it shows that the node has different equilibrium when parameters $U1, U2$ have different values and the results for $N=2, 3, 4, 5$ almost overlap. Meanwhile, it can be seen that the results of CDF also almost overlap. Therefore, we consider $N=2$ as the appropriate choice.

In Figure (\ref{fig8}), we show the effect of uncertain parameters on the resilience of whole network and each node. In Figure (\ref{fig8}), it is clear that when parameters are certain, network could maintain its resilience when average weight is greater than 7. However, with the effect of uncertain parameters, network could lose resilience even though its average weight is great than 7. With the growth of average weight, the network has more chance to be resilient. When the average weight is greater than a critical value, the network is absolute resilient. Similarly, in Figure (\ref{fig8}) (d) red part shows that when node's weight is greater than a critical value under certain average weight, the node could maintain its resilience. While, with the effect of uncertainty represented by blue part, a node may lose resilience even though its weight is greater than the critical value in Figure (\ref{fig4}) (b). Therefore, the method mentioned above could help us understand the effect of uncertainty on network-level and node-level resilience. Also, it help us to predict whether a node is resilient and the probability of a node to lose resilience.

\section{Conclusions}
At present, the research of how to estimate node-level resilience of dynamic networked system is still limited. Node level is important to make critical interventions to specific components whilst preserving our multi-scale understanding of general system behaviour. In this paper, an arbitrary polynomial chaos expansion (aPC) method is used to quantify the uncertainty of arbitrary uncertain distributions. This approach can effectively estimate node-level resilience and analyse the effect of uncertainty on each node. This could help us better make a prediction of the probability that a node loses its resilience and reduce the risk of uncertainty. In the future, we would like to survey how the community structure of network affects network-level and node-level resilience, for example, whether there exists a relationship between modularity of community in network and resilience.

\bibliographystyle{IEEEtran}
\bibliography{Ref}

\end{document}